\documentclass[proceedings]{rmaa}

\usepackage{rmaacite}

\renewcommand{\P}[1]{%
\ifnum#1=1\hbox{OW~168--326E}\fi
\ifnum#1=2\hbox{OW~167--317}\fi
\ifnum#1=3\hbox{OW~163--317}\fi
\ifnum#1=5\hbox{OW~158--323}\fi
\ifnum#1=0\hbox{OW~171--334}\fi}

\title{Explosions During Galaxy Formation}
\author{Hugo Martel and Paul R. Shapiro
  \affil{Dept. of Astronomy, University of Texas at Austin} }

\fulladdresses{
\item Hugo Martel and Paul R. Shapiro: Department of Astronomy,
  University of Texas, Austin, TX 78712, USA
  (hugo@simplicio.as.utexas.edu, shapiro@astro.as.utexas.edu).}

\shortauthor{Martel \& Shapiro}
\shorttitle{Explosions During Galaxy Formation}

\keywords{cosmology: theory --- galaxies: formation --- hydrodynamics --- 
  intergalactic medium}

\abstract{  
As an idealized model of the effects of energy release by supernovae 
during galaxy formation, we consider an explosion at the center of a
halo which forms at the intersection of filaments in the plane of
a cosmological pancake by gravitational instability during pancake
collapse. Such halos resemble the virialized objects found in
$N$-body simulations in a CDM universe and, therefore, serve as a convenient,
scale-free test-bed model for galaxy formation. $\rm ASPH/P^3M$ simulations
reveal that such explosions are anisotropic. The energy and metals
are channeled into the low density regions, away from the pancake plane. 
The pancake remains
essentially undisturbed, even if the explosion is strong enough
to blow away all the gas located inside the 
halo at the onset of the explosion and reheat the
IGM surrounding the pancake. 
Infall quickly replenishes this ejected gas 
and gradually restores the gas fraction as the halo mass continues
to grow. Estimates of the collapse epoch and  
SN energy-release for galaxies of different
mass in the CDM model can relate these results to scale-dependent
questions of blow-out and blow-away and their implication for 
early IGM heating and metal enrichment 
and the creation of dark-matter-dominated dwarf galaxies.
  }

\listofauthors{H.~Martel \& P.~R.~Shapiro}
\indexauthor{Martel, H.}
\indexauthor{Shapiro, P.~R.}

\begin{document}

\maketitle

\section{Introduction}
\label{sec:intro}
The release of energy that occurs during galaxy formation 
can have important consequences
for the structure and further evolution of these galaxies, other
galaxies, and the intergalactic medium (IGM) in which they form.
Numerous observations cannot be explained by theoretical models unless
energy release is invoked: e.g. 
(1) Observations reveal that dwarf spheroidal galaxies are dark-matter
rich relative to normal galaxies (e.g. Gallagher \& Wyse 1994). 
(2) Semi-analytical models of galaxy formation
in a CDM universe (e.g. White \& Frenk 1991) find that
gas cooling is too efficient,
leading to an overabundance of low-luminosity galaxies.  
(3) Simulations of galaxy formation fail to explain galactic rotation;
too much gas angular momentum is transferred to the
dark matter halo (e.g. Navarro \& Steinmetz 1997). (4)
$N$-body simulations of the CDM model predict an order of
magnitude more dwarf satellite galaxies in the Local Group than are observed
(e.g. Moore et al. 1999).
(5) Observational limits on the H I Gunn-Peterson effect
in the spectra of distant quasars
indicate that the IGM at $z\gtrsim5$
was already highly ionized (e.g. Songaila et al. 1999). 
(6) A heavy element abundance of $10^{-3}$ solar or more is ubiquitous
in the Ly$\alpha$ forest at $z\gtrsim3$, including that associated with gas
at close to the mean IGM density
(e.g. Ellison et al. 2000 and refs. therein).
(7) Observations of intracluster gas in X-ray clusters indicate a heavy
element abundance $\sim1/3$ solar and excess entropy relative
to the predictions of CDM simulations without energy-release
(e.g. Loewenstein \& Mushotzky 1996; Ponman, Cannon, \& Navarro 1999).

In this paper, we present 3D numerical 
gas dynamical simulations of the effect of energy release by
supernovae, and the consequences of this energy release for the
evolution of the halo in which the explosion takes place, the
surrounding large-scale structure of which the halo is a part, and
the IGM. Our first discussion of this work, based upon somewhat
lower-resolution simulations, was in Martel \& Shapiro (2000a).

\section{PANCAKE INSTABILITY AND FRAGMENTATION AS A TEST-BED FOR 
GALAXY FORMATION}

Galaxy formation which leads to star formation results in
supernova (SN) explosions
and the resulting shock-heating and outward acceleration of interstellar
and intergalactic gas.
Previous attempts to model this effect have typically been along one
of three lines, that which adopts a smooth initial gas distribution in a
galaxy-like, fixed, dark-matter gravitational potential well 
(e.g. Mac Low \& Ferrara 1999; Murakami \& Babul 1999;
Efstathiou 2000), 
that which considers a single, isolated, but evolving,
density fluctuation (i.e. without merging, infall, or external tidal
forces) (e.g. Katz 1992), and that
in which the galaxy forms by condensation out of Gaussian-random-noise
primordial density fluctuations such as in the CDM model
(e.g. Cen \& Ostriker 1992; Navarro \& White 1993; Gnedin \& Ostriker 1997;
Katz, Weinberg, \& Hernquist 1997; Steinmetz 1997; Yepes et al. 1997). 
In the first two approaches,
the computational ability to resolve shocks which propagate
away from the sites of explosive energy release is generally
higher, but because the galaxy is treated as isolated,
the cosmological initial and boundary conditions leading to the
formation of that galaxy are ignored. The third approach, which takes these
initial and boundary conditions into account, is generally more
realistic, but the resolving of shocks is generally quite poor.

\begin{figure}
  \begin{center}
    \leavevmode
   \includegraphics[width=\textwidth]{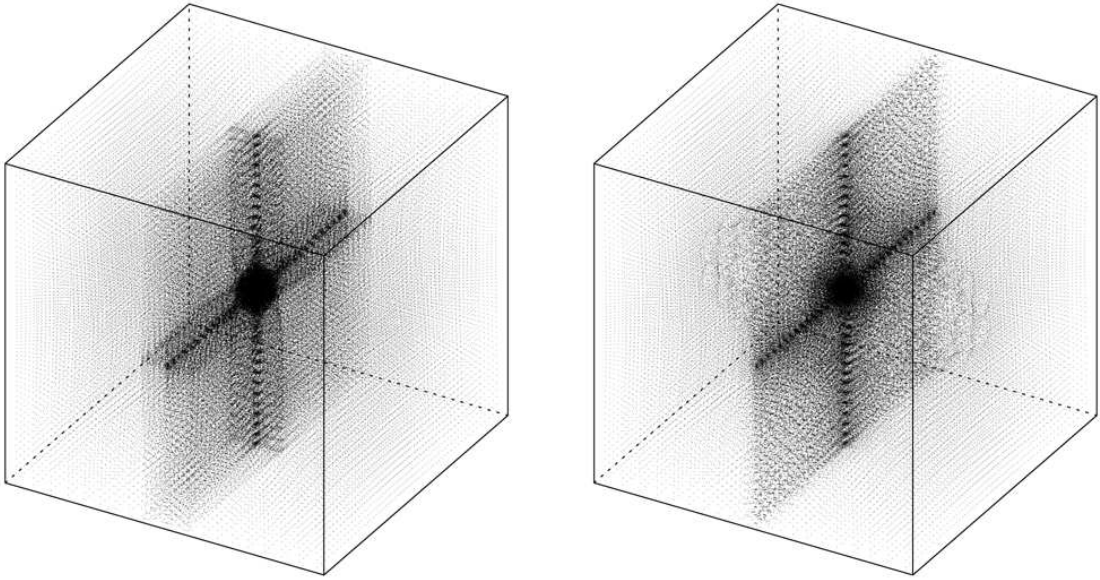}
    \caption{(a) Dark matter particles at $a/a_c=3.0$ for
             $\chi=0$; (b) gas particles at $a/a_c=3.0$ for
             $\chi=100$.}
    \label{fig:conf}
  \end{center}
\end{figure}


Structure 
formation from Gaussian random noise proceeds in a highly anisotropic way,
favoring the formation of pancakes and filaments over quasi-spherical
objects. Our previous work 
(Valinia et al. 1997;
Martel, Shapiro, \& Valinia 2000; Alvarez, Shapiro, \& Martel 2000)
has demonstrated, however, that a cosmological 
pancake, modeled as the nonlinear outcome of a single plane-wave density 
fluctuation, is subject to a linear gravitational instability which
results in the formation of
quasi-spherical lumps with density profiles similar to the universal halo
profile found to fit the results of 3D N-body simulations
in the CDM model (Navarro, Frenk, \& White 1997). This
suggests that this 3D instability
of cosmological pancakes may be used as an
alternative to the details of the
CDM model as a test-bed in which to study halo and galaxy formation
further.


This approach provides a good compromise between the two 
limits discussed above.
Since we are simulating a single galaxy, 
numerical resolution can 
be quite high, comparable to that of simulations 
which treat the galaxy as isolated. However, it provides a 
self-consistent cosmological origin and boundary condition for the galaxy,
including the important effects of anisotropic gravitational 
collapse and continuous infall, while avoiding the complexity of
simulations with Gaussian random noise.   

\section{SIMULATING THE EFFECT OF EXPLOSIVE ENERGY RELEASE ON GALAXY FORMATION}

Our gas dynamical simulations are based upon the 3D ASPH algorithm 
(Shapiro et al. 1996; Owen et al. 1998; Martel \& Shapiro 2000b),
coupled to a $\rm P^3M$ gravity solver, with
$64^3$ particles each of gas and dark matter, and
a $\rm P^3M$ grid of $128^3$ cells with softening length $\eta=0.3$ grid 
spacings. The initial conditions correspond to
the growing mode of a single sinusoidal plane-wave density
fluctuation, perturbed by two
transverse perturbation modes of equal wavelength and
amplitude equal to 1/5 of the amplitude of the primary pancake.
We model the release of energy due to
SNe in terms of a single impulsive explosion which may represent a starburst
or the collective effect of multiple SNe. The explosion occurs when
gas particles at the center of our dark matter
halo first reach a density contrast relative to the average background
density, $\rho_{\rm gas}/\langle\rho_{\rm gas}\rangle$, exceeding $10^3$
(at $a_{\rm exp}=1.912a_c$, where $a_c$
is the scale factor at which the primary pancake mode leads to caustic
formation in the dark matter and accretion shocks in the gas). We then
deposit a certain amount of thermal energy $\chi{\cal E}_{\rm halo}$
in the center of the halo, distributed smoothly over the central particles 
and their nearest neighbors, where ${\cal E}_{\rm halo}$ is the
total thermal energy of the gas
whose density exceeds 200 times the cosmic mean gas density,
and $\chi$ is a dimensionless constant. We have performed simulations for
$\chi=0$ (no explosion), $\chi=10$ (modest explosion),
$\chi=100$ (strong explosion) and $\chi=1000$
(blowaway regime). All simulations end at $a/a_c=3$. 

\begin{figure}
  \begin{center}
    \leavevmode
    \includegraphics[width=\textwidth]{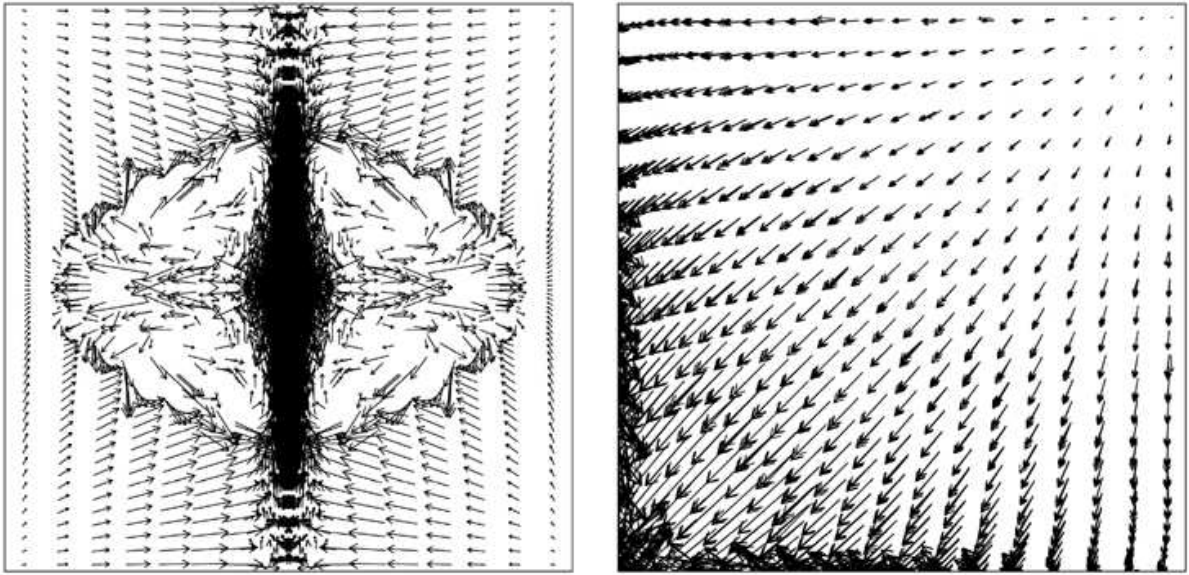}
    \caption{Velocity field of gas at $a/a_c=3.0$ for the case $\chi=100$:
             (left) in the plane normal to the pancake, which bisects the halo, 
             and (right) in the pancake plane.}
    \label{fig:vel}
  \end{center}
\end{figure}

We consider an Einstein-de Sitter universe
($\Omega_0=1$) with baryon density parameter $\Omega_{\rm B}\ll1$.
In that case, the
pancake problem is self-similar and scale-free (when radiative cooling
and photoheating are ignored), as long as all lengths are expressed
in units of the comoving pancake wavelength $\lambda_p$,
time is expressed in terms of $a/a_c$, and the explosion energy is
expressed in terms of the efficiency factor $\chi$.
This has the advantage that one simulation serves to model
all possible masses and collapse epochs, for each
value of $\chi$. Our neglect of radiative cooling,
a scale- and epoch-dependent process, amounts to the conservative
limiting approximation with which to assess the minimum
value of $\chi$ which leads to blow-away, metal-ejection, IGM shock-heating or 
the disturbance of the pancake. With cooling neglected, photoheating
can also be safely neglected, except for the possibility that a
photoheated background IGM (e.g. after reionization) might
have a significant enough pressure to influence the
outflow. We will evaluate the latter possibility in our discussion
section. Without heating and cooling, the assumption $\Omega_{\rm B}\ll1$
ensures that dark matter dominates the gravitational force everywhere
and that our results are essentially independent of $\Omega_{\rm B}$.
(Note: To be specific, we adopt $\Omega_{\rm B}=0.03$.)

\section{SCALE-FREE RESULTS}

Figure 1(a) shows the pancake-filament-halo
structure in the dark matter at $a/a_c=3$, for the case without explosion. 
The dark matter is hardly affected by the gas,
so Figure~1(a) represents, to very good accuracy, the final dark matter
distribution for all cases, with and without explosions.
Figure~1(b) shows the gas distribution for the intermediate case
$\chi=100$. The explosion expelled gas
from the central halo into the low-density regions surrounding the pancake,
sweeping out in the process exterior gas which was infalling along
directions perpendicular to the pancake plane. 
For $\chi=100$, the expelled gas fills an important fraction
of the computational volume at $a/a_c=3$, whereas for $\chi=1000$,
the expelled gas fills the entire volume outside the 
pancake-filament-halo structure.
In all cases with explosion, we found that
the pancake and filaments outside the halo are hardly affected. 

\begin{figure}
  \begin{center}
    \leavevmode
    \includegraphics[width=\textwidth]{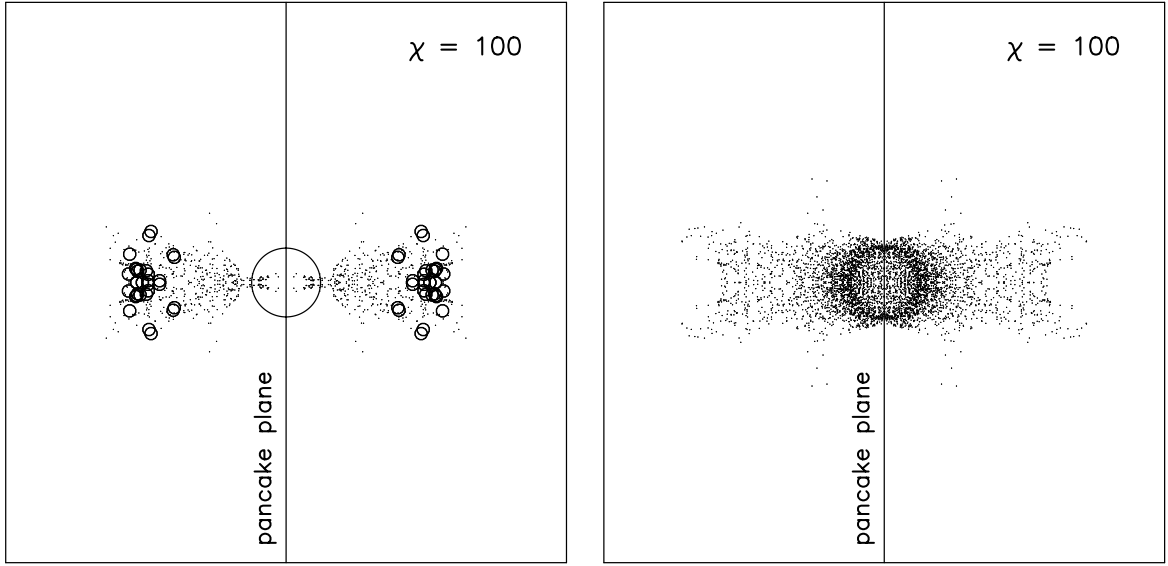}
    \caption{(Left) Gas particles at $a/a_c=3$ which were located inside
             the halo at onset of explosion (large circle radius $=r_{200}$
             at $a/a_c=3$). Open circles show particles
             that have been enriched in metals. (Right) 
             Gas particles located outside the halo at $a/a_c=3.0$, 
             that would have been inside if no explosion.}
    \label{fig:metals}
  \end{center}
\end{figure}

Figure 2 shows the velocity field of the gas at $a/a_c=3$
for the case $\chi=100$
in the plane normal to the pancake (left panel) and in
the pancake plane (right panel, only upper right quadrant 
plotted). The outer shock expanding into
the low density regions is clearly visible on the
left panel. The region bounded by this shock contains gas moving 
outward along the axis, forming a bipolar jet. The gas inside 
the pancake is still infalling toward the filaments and the central halo,
hardly affected by the explosion.
This infall will replenish the central halo with gas,
although some of this replenishment is subsequently
ejected in the bipolar jet. Neverthless, the halo eventually recovers
most of its gas fraction.


The effect of the explosion in blowing gas away is
illustrated by Figure~3.
Both panels show a projection on the plane normal to the
pancake at $a/a_c=3$, with the edge-on pancake central plane
and the halo represented by a vertical line and the large
circle of radius $r_{200}$,
respectively (where $r_{200}$ is the radius within which
the average total density is 200 times the cosmic mean). 
Gas particles which were located in the central 
halo (inside $r_{200}=0.0229$) at the onset of explosion are shown in
the left panel.
The small open circles show the subset of 128 particles which were the
original recipients of the explosion energy and, by implication,
the metal-enriched SN ejecta. 
By $a/a_c=3$, these particles have been ejected from the central
halo, but none into the pancake. The right panel
shows those gas particles which are located outside of the central halo 
at $a/a_c=3$ that would have been inside in the absence of explosion.
This is essentially the gas that was prevented from collapsing by the 
explosion. {\it Not a single particle that would
have been located in the halo in the absence of the explosion ends up
inside the surrounding pancake, away from the halo, instead}.


Without explosion, the total
halo mass $M_{\rm tot}$, the dark matter mass $M_{\rm dm}$
and gas mass $M_{\rm gas}$ within $r_{200}$
each grow by a factor of about 20 from the time of onset of the
explosion to $a/a_c=3$ (i.e. $a/a_{\rm exp}\cong1.5$). 
For the dark matter, the halo mass $M_{\rm dm}$
grows by this same factor with or without explosion.
This is an important difference between these simulations and those of
Mac Low \& Ferrara (1999), where explosions were triggered
inside a relaxed object of fixed mass. 

Our results indicate that the ability of explosions during galaxy
formation to eject gaseous baryons and metals is dependent on
the dimensionless energy-release efficiency $\chi$.
For a virialized halo at explosion onset, $\chi\gtrsim1$ is
in principle enough energy to ``unbind'' the gas inside the
halo. However, this assumes that the halo gas shares the
explosion energy equally; some may, instead, be ejected with more
energy than it needs, thereby depriving other gas of the minimum
it needs to escape. Even if the explosion energy is shared
equally, however, gas which escapes from the halo may not escape from the
gravitational pull of the surrounding pancake and filaments, and may, 
therefore, eventually fall back in. In a calculation like ours, with periodic, 
cosmological boundary conditions, it is not possible to identify a simple
escape velocity against which to compare our outflows to determine
what gas ultimately escapes, not only from its parent halo, but also from the
surrounding pancake. Nevertheless, we can distinguish the following
outcomes. For $\chi=10$, relatively little of the initial halo gas is blown 
out of the halo. The ejected fraction includes the gas in which the 
explosion energy and metals were initially deposited, but ejection is only
temporary, since it falls back in relatively quickly. For $\chi=100$, almost
3/4 of the halo gas is blown out, including that which received the original
explosion energy and metals. By $a/a_{\rm exp}\cong1.5$, it appears that the
metal-enriched gas is able to escape into the low-density valley between 
pancakes without falling back. Even so, infall along the pancake plane,
especially along the filaments, replenishes the ejected halo gas so
efficiently that by $a/a_{\rm exp}\cong1.05$, the halo contains more
gas than at the explosion onset, while by $a/a_{\rm exp}\cong1.5$,
it has an order of magnitude more gas than that, only 26\% less
than it would have had without explosion. Only for $\chi=1000$
is the halo gas all blown away and the explosion products driven so far
away so fast that they collide with their image gas expelled by the
explosion which occurred simultaneously in the neighboring pancake, when the
two explosion blast waves reach the same boundary from opposite
sides. Amazingly enough, even this ``blow-away'' explosion fails
to disturb the continuous 
equatorial rain of gas infalling from the surrounding
pancake plane and filaments. By $a/a_{\rm exp}\cong1.1$, the
expelled gas is fully replenished, while by
$a/a_{\rm exp}\cong1.5$, the final halo gas mass is close to 45\% of what
it would have been with no explosion.

In short, $\chi\lesssim10$ corresponds to the ``fall-back'' regime,
unlikely to be able to eject metals to pollute the IGM or
the surrounding pancake and filaments. Explosions with $\chi\sim100$ are
required for ``blow-out'' to occur, in which the gas which shared the 
original explosion energy and metals is ejected from the halo, but
not all the halo gas is ejected with it. A value of $\chi\sim1000$
is required for ``blow-away,'' in which all of the halo gas is expelled,
the explosion shock-heats not only the halo gas, but all of the external IGM,
as well, and the metals are thrown as far as the distance between
neighboring galaxies.

\section{APPLYING SCALE-FREE RESULTS TO SCALE-DEPENDENT GALAXY FORMATION}

All the results presented in \S 4 are scale-free. In this section, we
illustrate how we can apply 
these results to some particular scales of interest.
We use our pancake-halo model as a generic description of
structure formation at a particular length scale $\lambda=\lambda_p$.
To relate this model to the more complex
structures that form in the CDM model, we identify the redshift of 
pancake formation $z_c$ with the redshift $z_{\rm NL}$ at which density
fluctuations become nonlinear at the scale of $\lambda_p$. 
For an Einstein-de~Sitter universe, our pancake-halo model corresponds to a
``$\nu-\sigma$ density fluctuation'' if
$z_c=z_{\rm NL}=\nu\sigma_{\lambda_p}-1$, where $\sigma_{\lambda_p}$ is
the linearly-extrapolated present rms density fluctuation at scale $\lambda_p$.
We assume an untilted, cluster-normalized CDM model with 
$H_0=70\,\rm km\,s^{-1}Mpc^{-1}$ (i.e. $\sigma_8=0.53$).
For each value of $\lambda_p$ from 0.1
to $1\,\rm Mpc$ (comoving value, present units), we determine
$\sigma_{\lambda_p}$, $z_c$, the explosion onset $z_{\rm exp}$,
final redshift $z_f=(1+z_c)/3-1$, the
gas mass $M_{\rm gas}$ and total mass $M_{\rm tot}$
of the halo at $z_{\rm exp}$, the explosion
energy $E_{\rm exp}$ (i.e. $\chi{\cal E}_{\rm halo}$),
and the explosion efficiency $\epsilon\equiv 
E_{\rm exp}/M_{\rm gas}$. [Note: The comoving wavelength $\lambda_{\rm halo}$
which would encompass a mass equal to the halo mass at 
$a/a_c=a_{\rm exp}/a_c\cong2$ in the unperturbed background universe
is actually less than $\lambda_p$. Hence, in a CDM universe, fluctuations on 
the scale $\lambda_{\rm halo}$ on average grow to nonlinear amplitude
{\it earlier} than those on the scale $\lambda_p$. This effect is partially compensated here by the fact that the actual nonlinear collapse epoch
for a spherical fluctuation is later than that for a planar one of the same
initial amplitude. Neverthless, our approach here tends to underestimate 
the redshift of halo formation and explosion relative to the predictions
of, say, the Press-Schechter (``PS'') approximation for halos of the
same total mass. In the future, we will consider an alternative approach 
which more closely matches the epoch $a_c$ of our pancake collapse model to
the PS approximation. Our purpose here, however, is only to illustrate
how one would go about relating our scale-free determination of
the effects of a given explosion efficiency parameter $\chi$ to
the real efficiencies expected for galaxies of different mass and collapse
epochs. The net effect of a refinement of our prescription above for
identifying the values of $z_c$ to use for each $\lambda_p$ in
the CDM model will primarily be to make the results we describe below for
$3-\sigma$ fluctuations apply, instead, to ``$\nu-\sigma$''
fluctuations of lower $\nu$.]

We have considered a wide range of values for the explosion parameter $\chi$.
However, for any particular scale $\lambda_p$, 
we can use Milky Way star formation efficiencies and IMF to estimate
what a typical average 
value of $\chi$, $\chi_\lambda$, might be for that scale.
The efficiency of energy ejection is
$\epsilon=f_*\eta_{\rm SN}E_{\rm SN}$,
where $f_*$ is the star formation efficiency,
defined as the fraction of halo gas that turns into stars, $\eta_{\rm SN}$ is
the number of SNe expected per $M_\odot$ of stars formed,
and $E_{\rm SN}$ is the amount of energy per SN available
to drive the SN remnant blast wave after taking account of radiative
losses in the ISM. We take $f_*=0.05$ (Larson 1992;
Lada, Strom, \& Myers 1993; and refs. therein),
$\eta_{\rm SN}=5\times10^{-3}M_\odot^{-1}$ (i.e. 1 SN per $200M_\odot$ of stars)
(Murakami \& Babul 1999;
Efstathiou 2000; van den Bosch 2000), and $E_{\rm SN}=10^{50}\rm ergs$
(Thornton et al. 1998). 
Equating this estimated ``typical'' $\epsilon$
with our simulation efficiencies 
$\epsilon=\chi{\cal E}_{\rm halo}/M_{\rm gas}$ for
a given $z_c$ and $\lambda_p$,
we can then solve for $\chi_\lambda$. Our results for
$3-\sigma$ fluctuations are listed in 
Table~1, including the implied number of SNe per explosion, both
for any $\chi$ and also for $\chi=\chi_\lambda$.
As we see, our chosen range of explosion 
intensities, $\chi=10-1000$, covers all cases.

Our simulations show that $\chi\sim100$ is required to eject metals
into the IGM, while if $\chi\sim1000$,
the ejected gas and metals can travel all the way to the edge of the
computational volume by $a/a_{\rm exp}\cong1.5$. As Table~1 shows, 
for $3-\sigma$ fluctuations, a value of $\chi\sim100$ corresponds to
$\chi_\lambda$ only for $\lambda_p<0.2\,\rm Mpc$ (halo 
$M_{\rm tot}<10^7M_\odot$), while values as large as
$\chi\sim1000$ would require starburst efficiencies, well in excess of
current Milky Way star-formation estimates. Our analysis also allows us to
determine the metallicity yield implied by those explosions in such halos
and the implied space-averaged heavy element abundance of the IGM. We
will discuss that in a future paper. We note that our qualitatively new
result here, that even the most violent explosions which
blow gas away and eject metals nevertheless permit the parent halo to
replenish its lost gas in a fraction of a Hubble time, may enable
a single halo to contribute multiple outbursts.

\begin{table}
\caption{Illustrative Values for $3-\sigma$ Fluctuations in the SCDM Model}
\begin{center}
\begin{tabular}{cccccccrc}
\hline\hline
$\lambda_p$(Mpc) & $z_c$ & $z_{\rm exp}$ & $z_f$ &
$M_{\rm tot}(M_\odot)$ & 
$M_{\rm gas}(M_\odot)$ &
$N_{\rm SN}$ &
$\chi_\lambda$ &
$N_{\rm SN}(\chi=\chi_\lambda)$ \\ \hline
1.00 &  8.81 &  4.13 & 2.27     
& $1.37\times10^9$    & $3.36\times10^7$ 
& $2.95\times10^3\chi$    & 2.84 & 8380 \\
0.30 & 15.80 &  7.79 & 4.60     
& $3.71\times10^7$    & $9.08\times10^5$
& $1.23\times10^1\chi$   & 18.52 & 227 \\
0.20 & 18.56 &  9.23 & 5.52
& $1.10\times10^7$    & $2.70\times10^5$
& $1.88\times10^0\chi$     & 35.92 & 68 \\
0.10 & 23.66 & 11.90 & 7.22     
& $1.37\times10^6$    & $3.36\times10^4$
& $7.41\times10^{-2}\chi$  & 113.38 & 8 \\
\hline\hline
\end{tabular}
\end{center}
\end{table}

In the current scale-free simulations, the IGM in the background
universe is assumed to be cold enough that the actual pressure of the
IGM is negligible compared to that of the gas heated either by
the pancake accretion shocks or the explosion-generated shocks.
The evolution of the shocked pancake gas and the explosion, in that case,
are unaffected by the pressure of the IGM. As long as the pancake collapse
and the explosion took place before reionization of the IGM, this is a correct
description independent of the actual mass-scale of the pancake.
If reionization took place before the explosion, however, that
would have raised the pressure of the background IGM relative to that
in our current simulations.
Ferrara, Pettini, \& Shchekinov (2000) and
Murakami \& Babul (1999) suggest that the ejection of gas and
metals to large distances from early galaxies
would have been inhibited by the pressure of the IGM.
One concern, then, is that the efficiency of
metal ejection found in our simulations might be an artifact resulting 
from underestimating the IGM pressure. To check this, we compared the
total pressure $P_{\rm metals}$ of the ejecta (ram pressure 
$+$ thermal pressure)
to what the actual value of the IGM pressure, $P_{\rm IGM}$,
would have been if the IGM temperature following reionization were
$10^4\rm K$, for halos of different mass forming at different
epochs.
For $1-\sigma$ fluctuations, for
$\chi=100$, $P_{\rm metals}/P_{\rm IGM}>1$
for $\lambda_{\rm p}\geq0.60\rm\,Mpc$, while for $\chi=1000$, 
$P_{\rm metals}/P_{\rm IGM}>1$
for $\lambda_{\rm p}\geq0.20\rm\,Mpc$. For $3-\sigma$ fluctuations,
instead, for
$\chi=100$, $P_{\rm metals}/P_{\rm IGM}>1$
for $\lambda_{\rm p}\geq0.30\rm\,Mpc$, while for $\chi=1000$, 
$P_{\rm metals}/P_{\rm IGM}>1$
for $\lambda_{\rm p}\geq0.10\rm\,Mpc$. 
Our results are therefore valid
even if reionization occurred before the explosion,
except at small mass scales, where the IGM pressure might very well prevent
the ejection of metals over large distances.

All of the halos in Table~1 appear to form and
evolve early enough to cause widespread heavy element distribution
prior to $z=3$, as required to explain the ubiquitous metallicity of the
IGM measured in Lyman-$\alpha$ forest absorption spectra. The
smaller mass objects are the ones for which the smaller
binding energy per gram is more easily overcome by any given
star formation and supernova efficiency, if the latter are 
assumed to be independent of mass across the mass spectrum  of halos. 
Only the smallest mass objects in Table~1, in fact, can expect to do so
if they are limited to efficiencies typical for the Milky Way. Such small
mass halos are precisely the ones which are able to form early
enough that they may, indeed, do their exploding {\it before}
the reionization and reheating of the IGM is complete. If not, then
efficiencies much larger than
those of the current Milky Way are probably required,
if explosions from those objects are to succeed in blowing
the heavy elements into the IGM in the face of the opposing boundary pressure
of the IGM once the latter has been reionized.

A final determination of the success or
failure of heavy element distribution by SN explosions following
star formation in low-mass halos which form at high redshift,
we conclude, will
depend sensitively on the efficiencies of star formation and
supernova energy release, as well as the relative timing of these
explosions versus universal reionization, details which are highly
uncertain at this time. Our results support the conjecture that
heavy element distribution at the observed level of
$\approx10^{-3}$ solar in the IGM was accomplished
prior to the completion of reheating which accompanied universal reionization,
by a smaller fraction of the condensed baryons in the universe
than later were responsible for completing the reionization by
starlight or quasars.

\section{SUMMARY AND CONCLUSION}

We have simulated explosions inside cosmological halos which form
by gravitational instability during the collapse of a
cosmological pancake. This is a test-bed model that
can be used to describe explosions during galaxy formation under
more realistic circumstances, such as
that involving Gaussian random noise initial conditions. 
Our results include the following: 
(1) Blow-out and blow-away are generically 
anisotropic events which channel energy, mass loss, and
metal-enrichment outward 
preferentially along the symmetry axis of the local pancake and away from the
intersections of filaments in the pancake plane.
This anisotropy has a very different origin from that of the
blow-outs in Mac Low \& Ferrara (1999), which occurred in a 
rotationally-flattened gaseous disk inside an isolated, spherical halo
of dark matter, where the rotational-flattening introduced an
axisymmetry into the problem.
(2) Shock waves propagate into the IGM on both sides of
the pancake, heating the gas up to temperatures comparable to or greater than
the characteristic virial temperature of the halo which initiated the
explosion.
(3) Even with explosions which are
strong enough to propagate the shock waves all the way to the boundary 
of the computational volume, thereby filling the universe with overlapping
explosions which reheat the entire IGM, the condensed structure in which the 
explosion takes place is hardly disturbed.
(4) The explosion does not halt the continuous infall of gas from the 
surrounding pancake
and filaments into the halo. This infall replenishes the halo
gas and gradually restores its gas fraction.
As a result, the same halo may be able to experience multiple
explosions.

\acknowledgments

We are pleased to acknowledge stimulating discussions with Richard Mushotzky.
This work was supported by NASA Grants NAG5-7363 and
NAG5-7821, NSF Grants ASC-9504046, 
Grant 3658-0624-1999 from the Texas Advanced Research Program,
and the High Performance Computing Facility, University of Texas.



\begin{thebibliography}

\bibitem[x<1>]{x}
Alvarez, M., Shapiro, P. R., \& Martel, H. 2000, 
To appear in ``The Seventh Texas-Mexico Conference on Astrophysics:
  Flows, Blows, and Glows,'' eds. W. Lee and S. Torres-Peimbert, 
Rev.Mex.A.A. 
(Serie de Conferencias), in press (astro-ph/0006203)


\bibitem[x<1>]{x} 
Cen, R., \& Ostriker, J. P. 1992, ApJ, 399, L113



\bibitem[x<1>]{x}
Efstathiou, G. 2000, preprint (astro-ph/0002245)

\bibitem[x<1>]{x}
Ellison, S. L., Songaila, A., Schaye, J., \& Pettini, M. 2000, AJ, in press
(astro-ph/0005448)

\bibitem[x<1>]{x}
Ferrara, A., Pettini, M., \& Shchekinov, Y. 2000, preprint
(astro-ph/0004349)

\bibitem[x<1>]{x}
Gallagher, J. S., \& Wyse, R. F. G. 1994, PASP, 106, 706

\bibitem[x<1>]{x} 
Gnedin, N. Y., \& Ostriker, J. P. 1997, ApJ, 486, 581


\bibitem[x<1>]{x} 
Katz, N. 1992, ApJ, 391, 502

\bibitem[x<1>]{x} 
Katz, N., Weinberg, D. H., \& Hernquist, L. 1997,
ApJS, 105, 19

\bibitem[x<1>]{x}
Lada, E. A., Strom, K. M., \& Myers, P. C. 1993, in
Protostars and Planets III, eds. E. Levy \& J. I. Lunine
(University of Arizona Press), p. 245


\bibitem[x<1>]{x}
Larson, R. B. 1992, in Star Formation in Stellar Systems
(III Canary Islands Winter School of Astrophysics), eds.
G. Tenorio-Tagle, M. Prieto, \&
F. S\'anchez (Cambridge University Press), p. 127.

\bibitem[x<1>]{x}
Loewenstein, M., \& Mushotzky, R. F. 1996, ApJ, 466, 695

\bibitem[x<1>]{x} 
Mac Low, M.-M., \& Ferrara, A. 1999, ApJ, 513, 142

\bibitem[x<1>]{x} 
Martel, H., \& Shapiro, P. R. 2000a, Nucl. Phys. B, 80, 09/11
(astro-ph/9904121)

\bibitem[x<1>]{x} 
Martel, H., \& Shapiro, P. R. 2000b, in preparation

\bibitem[x<1>]{x} 
Martel, H., Shapiro, P. R., \& Valinia, A. 2000, in preparation

\bibitem[x<1>]{x}
Moore, B., Ghigna, S., Governato, F., Lake, G., Quinn, T., Stadel, J.,
\& Tozzi, P. 1999, ApJ, 524, L19

\bibitem[x<1>]{x}
Murakami, I., \& Babul, A. 1999, MNRAS, 309, 161

\bibitem[x<1>]{x} 
Navarro, J. F., Frenk, C. S., \& White, S. D. M. 1997, ApJ, 490, 493

\bibitem[x<1>]{x}
Navarro, J. F., \& Steinmetz, M. 1997, ApJ, 478, 13

\bibitem[x<1>]{x}
Navarro, J. F., \& White, S. D. M. 1993, MNRAS, 265, 271

\bibitem[x<1>]{x} 
Owen, J. M., Villumsen, J. V., Shapiro, P. R., \& Martel, H. 1998, 
ApJS, 116, 155

\bibitem[x<1>]{x}
Ponman, T. J., Cannon, D. B., \& Navarro, J. F. 1999, Nature, 397, 135



\bibitem[x<1>]{x} 
Shapiro, P. R., Martel, H., Villumsen, J. V., \& Owen, J. M. 
1996, ApJS, 103, 239

\bibitem[x<1>]{x}
Songaila, A., Hu, E. M., Cowie, L. L., \& McMahon, R. G. 1999,
ApJ, 525, L5

\bibitem[x<1>]{x} 
Steinmetz, M. 1997, in Structure and Evolution of the
Intergalactic Medium from QSO Absorption Line Systems, eds. P. Petitjean
and S. Charlot (Paris: Editions Fronti\`eres), p. 281

\bibitem[x<1>]{x}
Thornton, K., Gaudlitz, M., Janka, H.-Th., \& Steinmetz, M. 1998, ApJ, 500, 95



\bibitem[x<1>]{x} 
Valinia, A., Shapiro, P. R., Martel, H., \& Vishniac, E. T. 1997, ApJ, 479, 46

\bibitem[x<1>]{x}
van der Bosch, F. C. 2000, ApJ, 530, 177

\bibitem[x<1>]{x}
White, S. D. M., \& Frenk, C. S. 1991, ApJ, 379, 52

\bibitem[x<1>]{x} 
Yepes, G., Kates, R., Khokhlov, A., \& Klypin, A. 1997, MNRAS, 284, 235

\end{thebibliography}
\end{document}